\documentclass{article}
\usepackage{graphicx}
\usepackage{amsmath}

\input{tcilatex}

\begin{document}

\title{\textbf{Relativistic Particles and Commutator Algebras}}
\author{W. F. Chagas-Filho \\
Departamento de Fisica, Universidade Federal de Sergipe\\
Aracaju, SE, Brazil}
\maketitle

\begin{abstract}
We construct the vector fields associated to the space-time invariances of
relativistic particle theory in flat Euclidean space-time. We show that the
vector fields associated to the massive theory give rise to a differential
operator realization of the Poincare algebra, while the vector fields
associated to the massless theory, including the space-time supersymmetric
sector, allow extensions of the conformal algebra in terms of commutators.
\end{abstract}

\section{Introduction}

\noindent \noindent The relativistic particle continues to be one of the
most interesting dynamical systems to investigate if one wishes to try to
understand aspects of physics at a fundamental level. Among the many reasons
for this, we point out that relativistic particle theory has many features
that have higher-dimensional analogues in relativistic string theory while,
at the same time, is a prototype of general relativity. In the past few time
there has been an intense activity in trying to explain a small positive
measured value for the cosmological constant [1]. As a consequence of this
activity, a number of quite different and interesting analyses of the
cosmological constant problem have appeared [2]. In this context,
relativistic particle theory may be used as the simplest possible model for
the study of the cosmological constant problem because in the ``einbein''
version the relativistic particle action defines a one-dimensional generally
covariant field theory in which the particle's mass may be viewed as a
one-dimensional cosmological constant.

On the other side, one of the most interesting aspects of particle theory is
its connection with string theory. In the massless limit, the bosonic
particle action has a space-time supersymmetric extension [3] that exhibits
a local fermionic symmetry [4] which is crucial for the consistency of the
on-shell theory because it induces a perfect matching between the bosonic
and fermionic degrees of freedom. This space-time supersymmetric massless
particle action is then dimensionally extended to give rise to the
Green-Schwarz superstring [3]. The superstring action can then be further
extended to supermembranes [5]. However, while in particle theory the
existence of this local fermionic symmetry is a natural consequence of the
particle's own dynamics, in superstring and supermembrane theories
additional Wess-Zumino terms must be included in the action for the
fermionic symmetry to be present. This procedure of including Wess-Zumino
terms is thus quite artificial and so the origin of this fundamental
fermionic invariance, as a manifestation of the system's own dynamics, can
be appreciated only in the case of particle theory. This is one of the
interests of this work.

Another interesting point of particle theory, that attracted very little
attention, is the question if there exists a specific dimension for the
consistency of the supersymmetric theory. It is well known that quantum
superstrings seem to work only in space-time dimension $D=10$ while for
supermembranes the critical dimension seems to be $D=11$. However, no
restriction on the space-time dimension appears to exist in the case of the
superparticle. To parallel the treatment of superstrings, superparticle.
theories are then usually formulated in $D=10$. In this work we study this
problem at the classical level only but we are able to give evidence that
superparticle theory is a consistent theory only in $D=9$.

We are also interested here, as a way to gain some insight into the
cosmological constant problem, in the relationship between the presence of a
non-vanishing mass value in the bosonic particle action and the space-time
invariances of the action. We study this problem by associating a space-time
vector field to each of the space-time invariances of the action and then
computing the algebra defined by the generators of these vector fields. The
result we find is that the massless particle action has a larger set of
space-time invariances and then the appearance of a non-vanishing mass may
be associated to the breaking of some of these extra invariances. A positive
non-vanishing value for the cosmological constant could then also be the
result of symmetry-breaking mechanisms.

The paper is divided as follows: In section two we review the concept of
space-time vector fields and show how the generators of certain specific
vector fields give differential operator realizations of the Poincar\'{e}
algebra and of the conformal algebra in $D$ bigger than 2. In section three
we construct the vector field associated to the invariances of the massive
particle action in flat space-time and find that its generators give a
differential operator realization of the Poincar\'{e} algebra. Section four
deals with the massless particle action. We find that its invariances lead
to a differential operator realization of the conformal algebra. In
particular, we show that when the condition for free motion is satisfied, a
new invariance of the action permits the construction of an extension of the
conformal algebra. In section five we show how the features found in bosonic
particle theory may be extended to relativistic particles with space-time
supersymmetry. We present our conclusions in section six.

\section{Space-time vector fields}

Consider a Riemannian manifold with a metric of Euclidean signature 
\begin{equation}
ds^{2}=g_{\mu \nu }(x)dx^{\mu }dx^{\nu }  \tag{1}
\end{equation}
Under an infinitesimal coordinate transformation 
\begin{equation}
x^{\mu }\rightarrow x^{\mu }+\delta x^{\mu }  \tag{2}
\end{equation}
we have 
\begin{equation}
\delta g_{\mu \nu }=-(D_{\mu }\delta x_{\nu }+D_{\nu }\delta x_{\mu }) 
\tag{3}
\end{equation}
where 
\begin{equation}
D_{\mu }\delta x_{\nu }=\partial _{\mu }\delta x_{\nu }-\Gamma _{\mu \nu
}^{\lambda }\delta x_{\lambda }  \tag{4}
\end{equation}
and 
\begin{equation}
\Gamma _{\mu \nu }^{\lambda }=\frac{1}{2}g^{\lambda \delta }(\partial _{\mu
}g_{\delta \nu }+\partial _{\nu }g_{\mu \delta }-\partial _{\delta }g_{\mu
\nu })  \tag{5}
\end{equation}

Now let us consider the vector field 
\begin{equation}
\hat{R}(\xi )=\xi ^{\mu }\partial _{\mu }  \tag{6}
\end{equation}
such that 
\begin{equation}
D_{\mu }\xi _{\nu }+D_{\nu }\xi _{\mu }=0  \tag{7}
\end{equation}
The vector field $\hat{R}(\xi )$ gives rise to a coordinate transformation 
\begin{equation}
\delta x^{\mu }=\hat{R}(\xi )x^{\mu }=\xi ^{\mu }  \tag{8}
\end{equation}
under which 
\begin{equation}
\delta g_{\mu \nu }=-(D_{\mu }\xi _{\nu }+D_{\nu }\xi _{\mu })=0  \tag{9}
\end{equation}
The vector field $\hat{R}(\xi )$ is known as the Killing vector field and
generates isometries of the manifold. $\xi ^{\mu }$ is known as the Killing
vector. By definition, isometries are transformations which leave the metric
invariant and consequently correspond to transformations preserving the
length. The algebra of the Killing vector fields generates the algebra of
such symmetry transformations.

Now we restrict the discussion to the two special cases that are necessary
to understand the results of this work. The first case is that of Euclidean
flat space-time. This case is defined by the equations 
\begin{equation}
g_{\mu \nu }=\delta _{\mu \nu }  \tag{10a}
\end{equation}
\begin{equation}
\Gamma _{\mu \nu }^{\lambda }=0  \tag{10b}
\end{equation}
The Killing equation in this case becomes 
\begin{equation}
\partial _{\mu }\xi _{\nu }+\partial _{\nu }\xi _{\mu }=0  \tag{11}
\end{equation}
the solution of which is [6] 
\begin{equation}
\delta x_{\mu }=\xi _{\mu }=a_{\mu }+\omega _{\mu \nu }x^{\nu }  \tag{12}
\end{equation}
where 
\begin{equation}
a_{\mu }=cons\tan t  \tag{13a}
\end{equation}
\begin{equation}
\omega _{\mu \nu }=-\omega _{\nu \mu }=cons\tan t  \tag{13b}
\end{equation}
The Killing vector field in the case of a flat space-time then takes the
form 
\begin{eqnarray}
\hat{R}(\xi ) &=&\xi ^{\mu }\partial _{\mu }  \notag \\
&=&a^{\mu }\partial _{\mu }+\omega ^{\mu \nu }x_{\nu }\partial _{\mu } 
\notag \\
&=&a^{\mu }P_{\mu }-\frac{1}{2}\omega ^{\mu \nu }M_{\mu \nu }  \TCItag{14}
\end{eqnarray}
where we have defined 
\begin{equation}
P_{\mu }=\partial _{\mu }  \tag{15a}
\end{equation}
\begin{equation}
M_{\mu \nu }=x_{\mu }\partial _{\nu }-x_{\nu }\partial _{\mu }  \tag{15b}
\end{equation}
$P_{\mu }$ generates space-time translations and $M_{\mu \nu }$ generates
space-time rotations. The algebra of these generators is given by 
\begin{equation}
\lbrack P_{\mu },P_{\nu }]=0  \tag{16}
\end{equation}
\begin{equation}
\lbrack P_{\mu },M_{\nu \lambda }]=\delta _{\mu \nu }P_{\lambda }-\delta
_{\mu \lambda }P_{\nu }  \tag{17}
\end{equation}
\begin{equation}
\lbrack M_{\mu \nu },M_{\lambda \rho }]=\delta _{\nu \lambda }M_{\mu \rho
}+\delta _{\mu \rho }M_{\nu \lambda }-\delta _{\nu \rho }M_{\mu \lambda
}-\delta _{\mu \lambda }M_{\nu \rho }  \tag{18}
\end{equation}
where the brackets denote commutators. This is the Poincar\'{e} space-time
algebra in a commutator version and so the Killing field provides a
differential operator realization of the generators of the Poincar\'{e}
algebra.

The second case of interest here is the case of the conformal Killing
vectors. Consider now the vector field 
\begin{equation}
\hat{R}(\epsilon )=\epsilon ^{\mu }\partial _{\mu }  \tag{19}
\end{equation}
such that 
\begin{equation}
D_{\mu }\epsilon _{\nu }+D_{\nu }\epsilon _{\mu }=\frac{2}{d}g_{\mu \nu
}D.\epsilon  \tag{20}
\end{equation}
Here $d$ is the number of space-time dimensions and the factor $\frac{2}{d}$
on the right is needed for consistency.

For the flat space-time defined by equations (10) the conformal Killing
equation (20) becomes 
\begin{equation}
\partial _{\mu }\epsilon _{\nu }+\partial _{\nu }\epsilon _{\mu }=\frac{2}{d}%
\delta _{\mu \nu }\partial .\epsilon  \tag{21}
\end{equation}
and one can show that the most general solution for this equation for $d$
greater than $2$ is [6] 
\begin{equation}
\delta x^{\mu }=\epsilon ^{\mu }=a^{\mu }+\omega ^{\mu \nu }x_{\nu }+\lambda
x^{\mu }+(2x^{\mu }x^{\nu }-\delta ^{\mu \nu }x^{2})b_{\nu }  \tag{22}
\end{equation}
The vector field $\hat{R}(\epsilon )$ for the solution (22) is then given by 
\begin{equation}
\hat{R}(\epsilon )=a^{\mu }P_{\mu }-\frac{1}{2}\omega ^{\mu \nu }M_{\mu \nu
}+\lambda D+b^{\mu }K_{\mu }  \tag{23}
\end{equation}
where 
\begin{equation}
D=x^{\mu }\partial _{\mu }  \tag{24}
\end{equation}
and 
\begin{equation}
K_{\mu }=(2x_{\mu }x^{\nu }-\delta _{\mu }^{\nu }x^{2})\partial _{\nu } 
\tag{25}
\end{equation}
The two new additional vector fields (24) and (25) on the right of equation
(23) are respectively associated with dilatations and conformal boosts. The
generators of the vector field $\hat{R}(\epsilon )$ obey the commutator
algebra 
\begin{equation}
\lbrack P_{\mu },P_{\nu }]=0  \tag{26a}
\end{equation}
\begin{equation}
\lbrack P_{\mu },M_{\nu \lambda }]=\delta _{\mu \nu }P_{\lambda }-\delta
_{\mu \lambda }P_{\nu }  \tag{26b}
\end{equation}
\begin{equation}
\lbrack M_{\mu \nu },M_{\lambda \rho }]=\delta _{\nu \lambda }M_{\mu \rho
}+\delta _{\mu \rho }M_{\nu \lambda }-\delta _{\nu \rho }M_{\mu \lambda
}-\delta _{\mu \lambda }M_{\nu \rho }  \tag{26c}
\end{equation}
\begin{equation}
\lbrack D,D]=0  \tag{26d}
\end{equation}
\begin{equation}
\lbrack D,P_{\mu }]=-P_{\mu }  \tag{26e}
\end{equation}
\begin{equation}
\lbrack D,M_{\mu \nu }]=0  \tag{26f}
\end{equation}
\begin{equation}
\lbrack D,K_{\mu }]=K_{\mu }  \tag{26g}
\end{equation}
\begin{equation}
\lbrack P_{\mu },K_{\nu }]=2(\delta _{\mu \nu }D-M_{\mu \nu })  \tag{26h}
\end{equation}
\begin{equation}
\lbrack M_{\mu \nu },K_{\lambda }]=\delta _{\nu \lambda }K_{\mu }-\delta
_{\lambda \mu }K_{\nu }  \tag{26i}
\end{equation}
\begin{equation}
\lbrack K_{\mu },K_{\nu }]=0  \tag{26j}
\end{equation}
This is the differential operator realization of the conformal algebra in $d$
greater than 2, the extension of the Poincar\'{e} algebra (16-18).

\section{Relativistic particles}

The ``einbein'' action for a relativistic particle of mass $m$ in a flat $D$%
-dimensional space-time is given by [3] 
\begin{equation}
S=\frac{1}{2}\int d\tau (e^{-1}\dot{x}^{2}-em^{2})  \tag{27}
\end{equation}
where $\tau $ is an arbitrary parameter along the particle's world-line and
a dot denotes derivatives with respect to $\tau .$ The auxiliary coordinate $%
e(\tau )$ can be identified as the square root of a one-dimensional metric
[3] and so action (27) defines a generally covariant one-dimensional field
theory where the particle mass $m$ plays the role of a cosmological constant
[7]. The great advantage of action (27) is that it has a smooth transition
to the $m=0$ limit.

The classical equations of motion that follow from Hamilton's principle
applied to action (27) are 
\begin{equation}
\frac{1}{2}e^{-2}\dot{x}^{2}=-\frac{1}{2}m^{2}  \tag{28}
\end{equation}
\begin{equation}
\frac{d}{d\tau }(e^{-1}\dot{x}^{\mu })=0  \tag{29}
\end{equation}
Equation (29) shows that the particle will obey the free motion equation 
\begin{equation}
\ddot{x}^{\mu }=0  \tag{30}
\end{equation}
only when the condition $\frac{de}{d\tau }=0$ is satisfied. As the variable $%
e(\tau )$ may be associated to the geometry of the particle world-line, we
may interpret this condition as the condition for a constant, non-dynamical,
geometry. The canonical momentum conjugate to $x^{\mu }$ is given by 
\begin{equation}
p_{\mu }=e^{-1}\dot{x}_{\mu }  \tag{31}
\end{equation}
and so the equation of motion (28) gives rise to the Hamiltonian constraint
[8] 
\begin{equation}
\phi =\frac{1}{2}(p^{2}+m^{2})\approx 0  \tag{32}
\end{equation}
In this work we use Dirac's convention that a Hamiltonian constraint is set
equal to zero only after all calculations have been performed. In this sense
equation (32) means that $\phi $ ``weakly'' vanishes. The presence of the
first-class [8] constraint $\phi $ means that one of the $D$ \ $x^{\mu }$ is
an unobservable coordinate. Only $D-1$ of the $D$ \ $x^{\mu }$ are physical
coordinates.

Let us study the space-time invariances of action (27). It is invariant
under Poincar\'{e} transformations 
\begin{equation}
\delta x^{\mu }=a^{\mu }+\omega _{\nu }^{\mu }x^{\nu }  \tag{33a}
\end{equation}
\begin{equation}
\delta e=0  \tag{33b}
\end{equation}
and under the diffeomorphisms 
\begin{equation}
\delta x^{\mu }=\dot{\epsilon}x^{\mu }  \tag{34a}
\end{equation}
\begin{equation}
\delta e=\frac{d}{d\tau }(\epsilon e)  \tag{34b}
\end{equation}
In consequence of the invariance of action (27) under the Poincar\'{e}
transformation (33) we know that the following vector field can be defined
on space-time 
\begin{equation}
V=a^{\mu }P_{\mu }-\frac{1}{2}\omega ^{\mu \nu }M_{\mu \nu }  \tag{35}
\end{equation}
As we saw in the previous section, the generators of this field provide a
differential operator realization of the Poincar\'{e} algebra (16-18).

\section{Massless relativistic particles}

Let us now consider the massless particle action 
\begin{equation}
S=\frac{1}{2}\int d\tau e^{-1}\dot{x}^{2}  \tag{36}
\end{equation}
which is the $m=0$ limit of action (27). This action is also invariant under
the Poincar\'{e} transformations (33) and under the diffeomorphisms (34).
The equations of motion that follow from (36) are 
\begin{equation}
\frac{1}{2}e^{-2}\dot{x}^{2}=0  \tag{37}
\end{equation}
\begin{equation}
\frac{d}{d\tau }(e^{-1}\dot{x}^{\mu })=0  \tag{38}
\end{equation}
and as equation (38) indicates, the massless particle will only be free,
with a motion described by eq. (30), if the condition $\frac{de}{d\tau }=0$
is satisfied. If we compute the canonical momentum conjugate to $x^{\mu }$
that follows from actin (36) we get 
\begin{equation}
p^{\mu }=e^{-1}\dot{x}^{\mu }  \tag{39}
\end{equation}
and so equation (37) is equivalent to the first-class Hamiltonian constraint 
\begin{equation}
\phi =\frac{1}{2}p^{2}\approx 0  \tag{40}
\end{equation}
Again, only $D-1$ of the $x^{\mu }$ are physical. As we will see in the next
section, the supersymmetric extension of constraint $\phi $ will not only
reduce the bosonic degrees of freedom but will also create the condition for
the appearance of the local fermionic symmetry.

The massless action (36) has a larger set of space-time invariances than the
massive action (27). It is also invariant under the scale transformation 
\begin{equation}
\delta x^{\mu }=\lambda x^{\mu }  \tag{41a}
\end{equation}
\begin{equation}
\delta e=2\lambda e  \tag{41b}
\end{equation}
where $\lambda $ is a constant, and under the conformal transformation 
\begin{equation}
\delta x^{\mu }=(2x^{\mu }x^{\nu }-\delta ^{\mu \nu }x^{2})b_{\nu } 
\tag{42a}
\end{equation}
\begin{equation}
\delta e=4ex.b  \tag{42b}
\end{equation}
where $b_{\mu }$ is a constant vector. As a consequence of the invariances
of the massless action we can define, according to equation (22), the
space-time vector field 
\begin{equation}
V_{0}=a^{\mu }P_{\mu }-\frac{1}{2}\omega ^{\mu \nu }M_{\mu \nu }+\lambda
D+b^{\mu }K_{\mu }  \tag{43}
\end{equation}
As we saw in section two, the generators of the vector field $V_{0\text{ \ }%
} $provide a differential operator realization of the conformal algebra.

Let us now restrict the analysis to the case of free motion. Using equation
(30), it can be verified that the massless particle action (36) is also
invariant under the transformation 
\begin{equation}
x^{\mu }\rightarrow \exp \{\frac{1}{3}\beta (\dot{x}^{2})\}x^{\mu } 
\tag{44a}
\end{equation}
\begin{equation}
e\rightarrow \exp \{\frac{2}{3}\beta (\dot{x}^{2})\}e  \tag{44b}
\end{equation}
where $\beta $ is an arbitrary function of $\dot{x}^{2}$. This symmetry is
interesting because it has a higher-dimensional extension in the tensionless
limit of bosonic string theory [9] and, as the results of the next section
indicate, it may also be present in the tensionless limit of superstring
theory. Just as the massless limit is the high-energy limit of particle
theory, the tensionless limit [10] is the high-energy limit [11] of string
theory. Under the conditions required for the appearance of invariance (44),
we may say that the Hamiltonian constraint $\phi $ of equation (40) is the
generator of this invariance.

Invariance of the massless action under transformation (44) means that
infinitesimally we can define the scale transformation 
\begin{equation}
\delta x^{\mu }=\lambda \beta (\dot{x}^{2})x^{\mu }  \tag{45}
\end{equation}
where $\lambda $ is the same constant that appears in equations (22) and
(23). These transformations then lead to the existence of a new type of
dilatations. These new dilatations manifest themselves in the fact that the
vector field $D$ of equation (24) can be changed according to 
\begin{eqnarray}
D &=&x^{\mu }\partial _{\mu }\rightarrow D^{\ast }=x^{\mu }\partial _{\mu
}+\beta (\dot{x}^{2})x^{\mu }\partial _{\mu }  \notag \\
&=&D+\beta D  \TCItag{46}
\end{eqnarray}
In fact, because all vector fields in equation (43) involve partial
derivatives with respect to $x^{\mu }$ , and $\beta $ is a function of $\dot{%
x}^{\mu }$ \ only, we can also introduce the generators 
\begin{equation}
P_{\mu }^{\ast }=P_{\mu }+\beta P_{\mu }  \tag{47}
\end{equation}
\begin{equation}
M_{\mu \nu }^{\ast }=M_{\mu \nu }+\beta M_{\mu \nu }  \tag{48}
\end{equation}
\begin{equation}
K_{\mu }^{\ast }=K_{\mu }+\beta K_{\mu }  \tag{49}
\end{equation}
and define a new vector field $V_{0}^{\ast }$ \ by 
\begin{equation}
V_{0}^{\ast }=a^{\mu }P_{\mu }^{\ast }-\frac{1}{2}\omega ^{\mu \nu }M_{\mu
\nu }^{\ast }+\alpha D^{\ast }+b^{\mu }K_{\mu }^{\ast }  \tag{50}
\end{equation}
The generators of this vector field obey the commutator algebra 
\begin{equation}
\lbrack P_{\mu }^{\ast },P_{\nu }^{\ast }]=0  \tag{51a}
\end{equation}
\begin{equation}
\lbrack P_{\mu }^{\ast },M_{\nu \lambda }^{\ast }]=(\delta _{\mu \nu
}P_{\lambda }^{\ast }-\delta _{\mu \lambda }P_{\nu }^{\ast })+\beta (\delta
_{\mu \nu }P_{\lambda }^{\ast }-\delta _{\mu \lambda }P_{\nu }^{\ast }) 
\tag{51b}
\end{equation}
\begin{eqnarray}
\lbrack M_{\mu \nu }^{\ast },M_{\lambda \rho }^{\ast }] &=&(\delta _{\nu
\lambda }M_{\mu \rho }^{\ast }+\delta _{\mu \rho }M_{\nu \lambda }^{\ast
}-\delta _{\nu \rho }M_{\mu \lambda }^{\ast }-\delta _{\mu \lambda }M_{\nu
\rho }^{\ast })  \notag \\
&&+\beta (\delta _{\nu \lambda }M_{\mu \rho }^{\ast }+\delta _{\mu \rho
}M_{\nu \lambda }^{\ast }-\delta _{\nu \rho }M_{\mu \lambda }^{\ast }-\delta
_{\mu \lambda }M_{\nu \rho }^{\ast })  \TCItag{51c}
\end{eqnarray}
\begin{equation}
\lbrack D^{\ast },D^{\ast }]=0  \tag{51d}
\end{equation}
\begin{equation}
\lbrack D^{\ast },P_{\mu }^{\ast }]=-P_{\mu }^{\ast }-\beta P_{\mu }^{\ast }
\tag{51e}
\end{equation}
\begin{equation}
\lbrack D^{\ast },M_{\mu \nu }^{\ast }]=0  \tag{51f}
\end{equation}
\begin{equation}
\lbrack D^{\ast },K_{\mu }^{\ast }]=K_{\mu }^{\ast }+\beta K_{\mu }^{\ast } 
\tag{51g}
\end{equation}
\begin{equation}
\lbrack P_{\mu }^{\ast },K_{\nu }^{\ast }]=2(\delta _{\mu \nu }D^{\ast
}-M_{\mu \nu }^{\ast })+2\beta (\delta _{\mu \nu }D^{\ast }-M_{\mu \nu
}^{\ast })  \tag{51h}
\end{equation}
\begin{equation}
\lbrack M_{\mu \nu }^{\ast },K_{\lambda }^{\ast }]=(\delta _{\lambda \nu
}K_{\mu }^{\ast }-\delta _{\lambda \mu }K_{\nu }^{\ast })+\beta (\delta
_{\lambda \nu }K_{\mu }^{\ast }-\delta _{\lambda \mu }K_{\nu }^{\ast }) 
\tag{51i}
\end{equation}
\begin{equation}
\lbrack K_{\mu }^{\ast },K_{\nu }^{\ast }]=0  \tag{51j}
\end{equation}
Notice that the vanishing brackets of the conformal algebra (26) are
preserved as vanishing in the above algebra, but the non-vanishing brackets
of the conformal algebra now have linear and quadratic contributions from
the arbitrary function $\beta (\dot{x}^{2})$. It is interesting to choose $%
\beta $ simply linear in $\dot{x}^{2}$ because then, if the classical
equation of motion for $e(\tau )$ that follows from the massless action (36)
is imposed, transformation (44) becomes the identity transformation. The
algebra (51) is then on-shell in $x^{\mu }$ but off-shell in $e$. The
algebra (51) is an extension of the conformal algebra (26).

\section{Superparticles}

Finally we consider the case of relativistic particles with space-time
supersymmetry, also called superparticles. The bosonic massless particle
action (36) has the space-time supersymmetric extension [3] 
\begin{equation}
S=\frac{1}{2}\int d\tau e^{-1}(\dot{x}^{\mu }-i\bar{\theta}\Gamma ^{\mu }%
\dot{\theta})^{2}  \tag{52}
\end{equation}
\bigskip Here we choose $\theta _{\alpha }$ to be a $32$-component spinor ($%
\alpha =1,2,...,32)$. $\Gamma ^{\mu }$ are appropriate Dirac matrices and $%
\bar{\theta}=\theta ^{\dagger }\Gamma ^{0}$ . As will become clear in the
following, a consistent supersymmetric theory, with an equal number of
on-shell physical bosons and fermions, can only be defined if the space-time
dimension is $D=9$. Appropriate spinor and $\Gamma $ matrix representations
in this dimension are known to exist [5].

Introducing an infinitesimal constant Grassmann parameter $\varepsilon $,
spinor of the same type as the $\theta $ coordinate, supersymmetry is
realized in space-time as invariance of action (52) under the transformation
[3] 
\begin{equation}
\delta x^{\mu }=i\bar{\varepsilon}\Gamma ^{\mu }\theta  \tag{53a}
\end{equation}
\begin{equation}
\delta \theta =\varepsilon  \tag{53b}
\end{equation}
\begin{equation}
\delta e=0  \tag{53c}
\end{equation}
The equations of motion that follow from action (52) are 
\begin{equation}
\frac{1}{2}e^{-2}(\dot{x}^{\mu }-i\bar{\theta}\Gamma ^{\mu }\dot{\theta}%
)^{2}=0  \tag{54}
\end{equation}
\begin{equation}
\frac{d}{d\tau }(e^{-1}Z^{\mu })=0  \tag{55}
\end{equation}
\begin{equation}
\frac{d}{d\tau }(e^{-1}\Gamma .Z\theta )=0  \tag{56}
\end{equation}
where, following reference [3], we introduced the supersymmetric variable $%
Z^{\mu }=\dot{x}^{\mu }-i\bar{\theta}\Gamma ^{\mu }\dot{\theta}$. From
equations (55) and (56) it is clear that the superparticle will obey the
free motion equations [3] 
\begin{equation}
\frac{dZ^{\mu }}{d\tau }=\ddot{x}^{\mu }-i\bar{\theta}\Gamma ^{\mu }\ddot{%
\theta}=0  \tag{57}
\end{equation}
\begin{equation}
\Gamma .Z\dot{\theta}=0  \tag{58}
\end{equation}
only when the condition $\frac{de}{d\tau }=0$ \ is satisfied.

If we compute the canonical momentum conjugate to $x^{\mu }$ \ that follows
from action (52) we get 
\begin{equation}
p^{\mu }=e^{-1}(\dot{x}^{\mu }-i\bar{\theta}\Gamma ^{\mu }\dot{\theta}) 
\tag{59}
\end{equation}
and so equation (54) gives the Hamiltonian constraint 
\begin{equation}
\phi =\frac{1}{2}p^{2}\approx 0  \tag{60}
\end{equation}
The matrix $\Gamma .Z$ satisfies [3] 
\begin{equation}
(\Gamma .Z)^{2}=-Z^{2}=-e^{2}p^{2}=0  \tag{61}
\end{equation}
and as a result half of the $\theta _{\alpha }$ components in equation (58)
are eliminated from the theory. Only $16$ independent components remain. Now
action (52) is invariant under the fermionic $\kappa $-transformation [4] 
\begin{equation}
\delta \theta =i\Gamma .Z\kappa  \tag{62a}
\end{equation}
\begin{equation}
\delta x^{\mu }=i\bar{\theta}\Gamma ^{\mu }\delta \theta  \tag{62b}
\end{equation}
\begin{equation}
\delta e=4e(\bar{\theta}\dot{)}\kappa  \tag{62c}
\end{equation}
where $\kappa (\tau )$ \ is a Grassmann parameter. The matrix $\Gamma .Z$ \
appears again in the transformation equation for $\theta $ and invariance of
the action under (62) means that half of the $\theta _{\alpha }$ components
can be gauged away from the theory. Thus only 8 independent components
remain. This is a consequence of the presence of the Hamiltonian constraint $%
\phi $ of equation (60) which, as we saw, also reduces the number of bosonic
degrees of freedom to be $D-1$. A matching between the bosonic and fermionic
degrees of freedom can then only happen for $D=9$. This matching of 8
physical degrees of freedom of each type in superparticle theory also occurs
in the $D=10$ superstring [3] and in the $D=11$ supermembrane [5]. These
three theories describe the same number of physical degrees of freedom.

The superparticle action (52) is also invariant under the local bosonic
transformation [3] 
\begin{equation}
\delta \theta =\chi \dot{\theta}  \tag{63a}
\end{equation}
\begin{equation}
\delta x^{\mu }=i\bar{\theta}\Gamma ^{\mu }\delta \theta  \tag{63b}
\end{equation}
\begin{equation}
\delta e=0  \tag{63c}
\end{equation}
where $\chi (\tau )$ is a scalar parameter. Invariances (62) and (63) are
the usual gauge invariances of the massless superparticle.

Now we notice that from the Poincare transformation (33a) we have 
\begin{equation}
\delta \dot{x}^{\mu }=\delta (\frac{dx^{\mu }}{d\tau })=\frac{d}{d\tau }%
(\delta x^{\mu })=\omega _{\nu }^{\mu }\dot{x}^{\nu }  \tag{64}
\end{equation}
and the covariance of the supersymmetric variable $Z^{\mu }$ under Poincar%
\'{e} transformations requires that 
\begin{equation}
\delta (-i\bar{\theta}\Gamma ^{\mu }\dot{\theta})=\omega _{\nu }^{\mu }(-i%
\bar{\theta}\Gamma ^{\nu }\dot{\theta})  \tag{65}
\end{equation}
since $-i\bar{\theta}\Gamma ^{\mu }\dot{\theta}$ must behave as a Lorentz
vector, in the same way as $\dot{x}^{\mu }$. We can then define the Lorentz
transformation 
\begin{equation}
\delta Z^{\mu }=\omega _{\nu }^{\mu }Z^{\nu }  \tag{66a}
\end{equation}
\begin{equation}
\delta e=0  \tag{66b}
\end{equation}
and check that action (52) is invariant under transformation (66). Now we
point out that the superparticle action (52) also exhibits invariance under
a supersymmetric scale transformation give by 
\begin{equation}
\delta Z^{\mu }=\lambda Z^{\mu }  \tag{67a}
\end{equation}
\begin{equation}
\delta e=2\lambda e  \tag{67b}
\end{equation}
and also under the supersymmetric conformal transformation 
\begin{equation}
\delta Z^{\mu }=(2Z^{\mu }Z^{\nu }-\delta ^{\mu \nu }Z^{2})B_{\nu } 
\tag{68a}
\end{equation}
\begin{equation}
\delta e=2Z.Be  \tag{68b}
\end{equation}
where $B_{\mu }$ is a constant vector, which can be chosen for instance as
the value of $\ Z_{\mu }$ when $\tau =0$. Our experience with the massless
particle leads us to expect that, as a consequence of the invariance of the
superparticle action (52) under transformations (66), (67) and (68), the
vector field 
\begin{equation}
\bar{V}_{0}=-\frac{1}{2}\omega ^{\mu \nu }\bar{M}_{\mu \nu }+\lambda \bar{D}%
+B^{\mu }\bar{K}_{\mu }  \tag{69}
\end{equation}
may be defined in space-time. The generators of this vector field are given
by 
\begin{equation}
\bar{M}_{\mu \nu }=Z_{\mu }\frac{\partial }{\partial Z^{\nu }}-Z_{\nu }\frac{%
\partial }{\partial Z^{\mu }}  \tag{70}
\end{equation}
\begin{equation}
\bar{D}=Z^{\mu }\frac{\partial }{\partial Z^{\mu }}  \tag{71}
\end{equation}
\begin{equation}
\bar{K}_{\mu }=(2Z_{\mu }Z^{\nu }-\delta _{\mu }^{\nu }Z^{2})\frac{\partial 
}{\partial Z^{\nu }}  \tag{72}
\end{equation}
These generators obey the algebra 
\begin{equation}
\lbrack \bar{M}_{\mu \nu },\bar{M}_{\lambda \rho }]=\delta _{\nu \lambda }%
\bar{M}_{\mu \rho }+\delta _{\mu \rho }\bar{M}_{\nu \lambda }-\delta _{\nu
\rho }\bar{M}_{\mu \lambda }-\delta _{\mu \lambda }\bar{M}_{\nu \rho } 
\tag{73a}
\end{equation}
\begin{equation}
\lbrack \bar{D},\bar{D}]=0  \tag{73b}
\end{equation}
\begin{equation}
\lbrack \bar{D},\bar{M}_{\mu \nu }]=0  \tag{73c}
\end{equation}
\begin{equation}
\lbrack \bar{D},\bar{K}_{\mu }]=\bar{K}_{\mu }  \tag{73d}
\end{equation}
\begin{equation}
\lbrack \bar{M}_{\mu \nu },\bar{K}_{\lambda }]=\delta _{\nu \lambda }\bar{K}%
_{\mu }-\delta _{\lambda \mu }K_{\nu }  \tag{73e}
\end{equation}
\begin{equation}
\lbrack \bar{K}_{\mu },\bar{K}_{\nu }]=0  \tag{73f}
\end{equation}
Because the supersymmetric action (52) is not invariant under translations
of the $Z^{\mu }$ variable, this algebra is the supersymmetric extension of
the bosonic conformal algebra (26).

To conclude this work we point out that when equation (57) for free motion
is satisfied the superparticle action (52) is invariant under the
supersymmetric extension of the bosonic massless particle transformation
(44). The transformation equations are 
\begin{equation}
x^{\mu }\rightarrow \exp \{\frac{1}{3}\gamma (Z^{2})\}x^{\mu }  \tag{74a}
\end{equation}
\begin{equation}
\theta \rightarrow \exp \{\frac{1}{6}\gamma (Z^{2})\}\theta   \tag{74b}
\end{equation}
\begin{equation}
e\rightarrow \exp \{\frac{2}{3}\gamma (Z^{2})\}e  \tag{74c}
\end{equation}
where $\gamma $ is an arbitrary function of $Z^{2}$. As a consequence of \
the invariance of action (52) under this transformation, we may extend the
scale transformations (67a) to transformations of the type 
\begin{equation}
\delta Z^{\mu }=\lambda \gamma (Z^{2})Z^{\mu }  \tag{75}
\end{equation}
which in turn leads to the existence of the generator 
\begin{equation}
\bar{D}^{\ast }=\gamma (Z^{2})\bar{D}=\gamma (Z^{2})Z^{\mu }\frac{\partial }{%
\partial Z^{\mu }}  \tag{76}
\end{equation}
Here we can not proceed as in the bosonic case and introduce also extended
operators to $\bar{M}_{\mu \nu }$ and $K_{\mu }$ because these operators
involve partial derivatives with respect to $Z^{\mu }$. The best we can do
is define the extended vector field 
\begin{equation}
\bar{V}_{0}^{\ast }=-\frac{1}{2}\omega ^{\mu \nu }\bar{M}_{\mu \nu }+\lambda
(\bar{D}+\bar{D}^{\ast })+B^{\mu }\bar{K}_{\mu }  \tag{77}
\end{equation}
and the generators of this vector field obey the algebra (73) complemented
with the equations 
\begin{equation}
\lbrack \bar{D},\bar{D}^{\ast }]=\bar{D}\gamma \bar{D}  \tag{78}
\end{equation}
\begin{equation}
\lbrack \bar{M}_{\mu \nu },\bar{D}^{\ast }]=\bar{M}_{\mu \nu }\gamma \bar{D}
\tag{79}
\end{equation}
\begin{equation}
\lbrack \bar{K}_{\mu },\gamma \bar{D}]=\bar{K}_{\mu }\gamma \bar{D}-\gamma 
\bar{K}_{\mu }  \tag{80}
\end{equation}

\section{Conclusions}

As the first step, in this work we reviewed the concept of space-time vector
fields and how, in the case of Euclidean flat space, these vector fields
furnish differential operator realizations of the Poincare and conformal
algebra when the space-time dimension is greater than two. Essentially
speaking, the origin of these vector fields are space-time transformations.
The space-time transformations that leave the relativistic particle action
invariant where then used to associate vector fields to the action, and the
algebra of the generators of these vector fields was displayed. We saw that
the invariances of the massive particle action give rise to a commutator
realization of the Poincar\'{e} algebra. Taking the limit when the
particle's mass goes to zero, we saw that the action becomes more symmetric,
being now also invariant under scale and conformal transformations. As a
consequence of these extra invariances a larger vector field can be defined
in space-time. The generators of this larger vector field were demonstrated
to realize a commutator version of the conformal algebra.

An interesting aspect of the free massless theory was shown to be the
existence of another type of scale transformation that mixes the dynamical
variable $x^{\mu }$ with its derivative with respect to the parameter used
to describe the dynamical evolution of the relativistic particle.
Transformations that mix the dynamical variables with their derivatives have
been classified by some authors as non-internal gauge transformations [12],
[13] and we may interpret transformation (44) as one such transformation in
the context of generally covariant systems. As we saw, transformation (44),
which is present only in the free theory, gives rise to an extension of the
conformal algebra.

If we treat the relativistic particle action as a toy model for the study of
the cosmological constant problem then the above results allow us to picture
the appearance of a particle mass as a sequence of symmetry-breaking
mechanisms that induce a transition from the extended conformal algebra (51)
down to the Poincar\'{e} algebra (16-18). The first mechanism must induce an
interaction. When an interaction appears invariance (44) breaks down and we
have a reduction from the extended algebra (51) to the smaller conformal
algebra (26). Breakdown of the conformal algebra (26) then reduces the
invariances of the action to Poincar\'{e} invariance, having as a
consequence the appearance of a non-vanishing mass.

As a final analyses in this work we verified if supersymmetric extensions of
the algebra (51) could be constructed in the case of massless free
superparticles. Although the extension is not straightforward because the
action is not invariant under translations of the supersymmetric variable $%
Z^{\mu }$, and so the theory is not Poincar\'{e} invariant but just Lorentz
invariant, we were able to construct a supersymmetric vector field whose
generators, defined in terms of the supersymmetric variable $Z^{\mu },$
partially realize a supersymmetric extension of the commutator algebra (51)
for the case of a free massless superparticle. The same symmetry-breaking
mechanisms we mentioned in the case of the massless bosonic particle may
then also be present in the supersymmetric version of the theory.

\end{document}